\documentclass[journal]{IEEEtran} 
\usepackage{cite}
\usepackage{amsmath,amssymb,amsfonts}
\usepackage{algorithm}
\usepackage{algorithmic}
\usepackage{graphicx}
\usepackage{epstopdf}
\usepackage{textcomp}
\usepackage{array}
\usepackage{multirow}
\usepackage{color}
\usepackage{url}
\usepackage{subfigure}
\usepackage{setspace}
\usepackage{hyperref}
\usepackage[hyphenbreaks]{breakurl}
\usepackage{threeparttable}
\usepackage{makecell}

\usepackage{booktabs}

\begin{document}
\title{SigNet: A Novel Deep Learning Framework for Radio Signal Classification}
\author{
    \vskip 1em
    Zhuangzhi~Chen*,
    Hui~Cui*,
    Jingyang~Xiang,
    Kunfeng~Qiu,
    Liang~Huang, \emph{Member,~IEEE},\\
    Shilian~Zheng, \emph{Member,~IEEE},
    Shichuan~Chen,
    Qi~Xuan, \emph{Senior Member,~IEEE},
    and~Xiaoniu~Yang
    \thanks{
    *: The first two authors contributed equally.

    This work was supported in part by the National Natural Science Foundation of China under Grant 61973273 and Grant 62072410, and by the Zhejiang Provincial Natural Science Foundation of China under Grant LR19F030001 and LY19F020033. \emph{(Corresponding author: Qi Xuan.)}

    Z. Chen, H. Cui, J. Xiang, and K. Qiu are with the Institute of Cyberspace Security, and also with the College of Information Engineering, Zhejiang University of Technology, Hangzhou 310023, China (e-mail: zhuangzhichen@foxmail.com; huicuizjut@qq.com; xiangxiangjingyang@gmail.com; yexijoe@163.com).
    
    L. Huang is with the College of Computer Science and Technology, Zhejiang University of Technology, Hangzhou, China (email: lianghuang@zjut.edu.cn)
    
    S. Zheng and S. Chen are with the Science and Technology on Communication Information Security Control Laboratory, Jiaxing 314033, China (e-mail: lianshizheng@126.com; sicanier@sina.com).
    
    Q. Xuan is with the Institute of Cyberspace Security, College of Information Engineering, Zhejiang University of Technology, Hangzhou 310023, China, with the PCL Research Center of Networks and Communications, Peng Cheng Laboratory, Shenzhen 518000, China, and also with the Utron Technology Co., Ltd. (as Hangzhou Qianjiang Distinguished Expert), Hangzhou 310056, China (e-mail: xuanqi@zjut.edu.cn).
    
    X. Yang is with the Institute of Cyberspace Security, Zhejiang University of Technology, Hangzhou 310023, China, and also with the Science and Technology on Communication Information Security Control Laboratory, Jiaxing 314033, China (e-mail: yxn2117@126.com).
    }
}
\maketitle
\begin{abstract}
Deep learning methods achieve great success in many areas due to their powerful feature extraction capabilities and end-to-end training mechanism, and recently they are also introduced for radio signal modulation classification. In this paper, we propose a novel deep learning framework called SigNet, where a signal-to-matrix (S2M) operator is adopted to convert the original signal into a square matrix first and is co-trained with a follow-up CNN architecture for classification. This model is further accelerated by integrating 1D convolution operators, leading to the upgraded model SigNet2.0. The simulations on two signal datasets show that both SigNet and SigNet2.0 outperform a number of well-known baselines. More interestingly, our proposed models behave extremely well in small-sample learning when only a small training dataset is provided. They can achieve a relatively high accuracy even when 1\% training data are kept, while other baseline models may lose their effectiveness much more quickly as the datasets get smaller. Such result suggests that SigNet/SigNet2.0 could be extremely useful in the situations where labeled signal data are difficult to obtain. The visualization of the output features of our models demonstrates that our model can well divide different modulation types of signals in the feature hyper-space.

\end{abstract}



\section{Introduction}
\label{sec:introduction}
\IEEEPARstart{Q}{uick} and accurate classification of radio signals has become an important issue in intelligent signal processing, due to the wide usage of radio technology in numerous fields. In the open electromagnetic space, there are various radio signal classification tasks according to different standards and applications, such as modulation recognition~\cite{8720067}, Aircraft Communications Addressing and Reporting System (ACARS) signal classification~\cite{8922798, 8750803}, technology recognition~\cite{8935690,8824856} etc. With the massive use of mobile devices and the development of 5G technology~\cite{gupta2015a}, the demand for limited electromagnetic spectrum resources in modern society is growing rapidly, which increases the difficulty of managing radio signals in open electromagnetic space. Improving the accuracy of modulation recognition can quickly manage the effective range of electromagnetic spectrum, and ensure the safety and reliability of communication system~\cite{zheng2018big}. Radio signal classification and modulation recognition have drawn the attention of many researchers. Recently, Kulin et al.~\cite{KulinEnd}, Mendis et al.~\cite{MendisDeep}, Dobre~\cite{Octavia2015Signal} and Zheng et al.~\cite{zheng2019deep} have proposed their respective radio signal recognition methods.

Signal classification has generally been accomplished by feature extraction and classification. Feature extraction focuses on extracting the important characteristics from the target signals, such as instantaneous amplitude, frequency and phase, constellation diagram, higher-order moment, and time-frequency diagram etc., which then are fed into machine learning methods for classification. For example, Walenczykowska et al.~\cite{walenczykowska2016type} used wavelet transform and neural network for modulation identification. Li et al.~\cite{li2019wavelet} mapped the signal to a high-dimensional feature space by using wavelet transform and then used Support Vector Machine (SVM) to automatically recognize multiple types of digitally modulated signals. Triantafyllakis et al.~\cite{triantafyllakis2017phasma} adopted random forest algorithm to identify pulse-time-frequency image characteristics of intra-pulse modulation. Vu{\v{c}}i{\'c} et al.~\cite{vuvcic2017cyclic} introduced a theory based on cyclic stationary signal classification method. While Abdelmutalab et al.~\cite{abdelmutalab2016automatic} used high order cumulants as features to classify M-PSK and M-QAM signals. Most of these methods require extensive expertise and meanwhile may be lack of high precision, which greatly limit their application, especially in complex scenes.

Recently, with the successful application of deep learning in image classification~\cite{xuan2017automatic,xuan2018multiview}, textural analysis~\cite{chevitarese2018seismic}, speech recognition~\cite{QianVery}, and graph mining~\cite{xuan2019subgraph,chen2019lstm}, deep neural networks are also becoming a preferable way for signal classification. Deep learning models, especially Convolutional Neural Networks (CNNs), can automatically extract features from various datasets based on task goals, which have been proved to be significantly superior to manual feature extraction in many situations. Due to their powerful feature learning capabilities, CNNs are widely applied for signal classification. For instance, O’Shea et al.~\cite{8267032} adopted VGG architecture principals to design a 1D-CNN model, making it suitable for small radio signal classification tasks; meanwhile they also adopted a $1\times{1}$ convolutional layer for ResNet unit input, achieving comparable results. O’Shea et al.~\cite{o2016convolutional} also used a narrow 2D-CNN for radio modulation recognition by simply considering both in-phase (I) and quadrature (Q) signal sequences in the time domain at the same time. A more complex model in their paper was called CNN2, based on which Liu et al.~\cite{liu2017deep} and West et al.~\cite{west2017deep} designed their own deep learning models for the respective signal modulation classification tasks. Using such kind of 1D convolution and narrow 2D convolution, Xu et al.~\cite{xu2020spatiotemporal} proposed a multi-channel CNN, named MCLDNN, to do modulation classification with different convolution features. However, these existing models have extremely simple structures and may not take advantage of the deep architecture of neural networks to deal with more complex scenes. And for narrow 2D CNN technique (using a narrow width input), it may not adequately utilize the power of CNNs, since when the pooling layers or stride $>1$, feature maps in hidden layers will shrink to one-dimensional vectors soon, and thus they can only extract features in this one direction.

An alternative way is to preprocess the signals, so as to change them into matrices, which are then fed into CNNs to get classification results. For instance, Peng et al.~\cite{peng2018modulation} proposed a modulation classification algorithm based on the signal constellation diagram and deep learning models. Zeng et al.~\cite{zeng2019spectrum} transformed signals into spectrogram images as the input of CNN. There are also some related algorithms in the area of time series analysis. For example, Wang et al.~\cite{wang2016imaging} proposed a novel framework for encoding time series, where the Gramian Angular Fields (GAF) and Markov Transition Fields (MTF) are used to convert time series into different types of images. However, these conversion methods are based on specific mathematical formulas, like some specific feature transformation (transform the original data to a more complex but specific space), and thus are lack of flexibility and may lose some useful features.


On the other hand, radio signals are also a kind of time-series, therefore, instead of losing their time information in CNNs, long short term memory (LSTM)~\cite{Greff2015LSTM} can be a good candidate to capture the time related features of signals. In fact,
LSTM has been adopted in multiple signal fields~\cite{yildirim2018novel, mostayed2018classification}. For instance, O’Shea et al.~\cite{o2016end} applied deep recurrent neural networks (RNN) for end-to-end radio traffic sequence recognition. Yildirim et al.~\cite{yildirim2018novel} proposed a deep bidirectional LSTM network for wavelet sequences and realized electrocardiogram signal classification. Mostayed et al.~\cite{mostayed2018classification} used a bidirectional LSTM network to detect pathologies in ECG signals. Rajendran et al.~\cite{rajendran2018deep} adopted an LSTM based model for modulation classification in a distributed wireless spectrum sensing network. Huang et al.~\cite{huang2019data} did three kinds of signal data augmentations and evaluated them through an LSTM model. LSTM models are able to capture the time information due to its gate structure and memory cells and thus are considered to be more suitable for signal classification, leading to higher accuracy in general. However, they are always time-consuming and are also difficult to train especially for long signals in complex tasks.

To fully utilize the powerful feature extraction capability of CNNs and meanwhile overcome the shortage of above methods, in this paper, we first focus on transforming signal to a two-dimensional matrix as input so as to make it easy for CNN to extract features. In particular, we design a flexible deep learning framework \emph{SigNet} based on sliding trainable operators which are used to adaptively extract the characteristics of signals. Since different operators can capture the signal properties in different aspects and scales and further automatically transfer a signal into different matrices, integration of which by typical CNN architecture thus can largely improve the classification performance. Moreover, to further accelerate the training of SigNet, we replace some of the 2D convolution operators by 1D convolution operators, and propose SigNet2.0. Interestingly, SigNet2.0 is not only faster than SigNet, but may also have slightly higher classification accuracy, indicating its outstanding performance in practice. The main contributions of this paper are as follows.

\begin{enumerate}
\item We develop SigNet as a novel deep learning framework for signal classification by introducing a sliding square operator S2M, with which the input signal can be automatically transformed into a square feature matrix. This S2M is trainable by connecting to a CNN structure and can adaptively extract important features, avoiding tedious preprocessing of the original signals. 

\item We accelerate SigNet by integrating 1D and 2D convolution operators, and propose SigNet2.0. This upgraded model reduces the complexity of signals through 1D convolution operators and then converts the simplified signals into feature matrices through S2M, thereby improving both classification accuracy and efficiency.

\item We perform comprehensive simulations on two signal datasets, and validate the outstanding performance of our models. Specifically, SigNet performs significantly better than other models, and is close to the latest model MCLDNN on RML2016.10a dataset (The accuracy of SigNet on the entire test set is 61.35\%, while that of MCLDNN is 61.16\%). And for dataset Sig2019-12, SigNet performs significantly better than all the other models, especilly when the SNR is below 0dB. (The accuracy of SigNet on the entire test set is 72.69\%, while the second best model MCLDNN is 67.52\%), SigNet2.0 performs even better on two datasets (achieving the accuracy of 62.30\% and 72.87\%, respectively). The visualizations of the output features for various models shows that our proposed models have a strong ability to distinguish features of modulation categories.

\item We find that SigNet and SigNet2.0 perform extremely well in small-sample learning. Both of them keep a relatively high accuracy when using a much smaller training dataset, while other methods may lose their effectiveness quickly. This phenomenon suggests that our proposed models could be more practical since labeling signals are always expensive.

\end{enumerate}

The rest of paper is organized as follows. In Sec.~\ref{sec:method}, we present the details of our technique and the principal and advantages. After that, we show the simulation setup and results with analysis and discussion in Sec.~\ref{sec:experiments}. And we further visualize the model output features in Sec.~\ref{sec:visual} to explain how SigNet and SigNet2.0 behave better than other models. Finally, we conclude the paper and give possible directions for the future work in Sec.~\ref{sec:conclusion}.

\begin{figure*} [!t]
\centering
\includegraphics[width=0.99\textwidth]{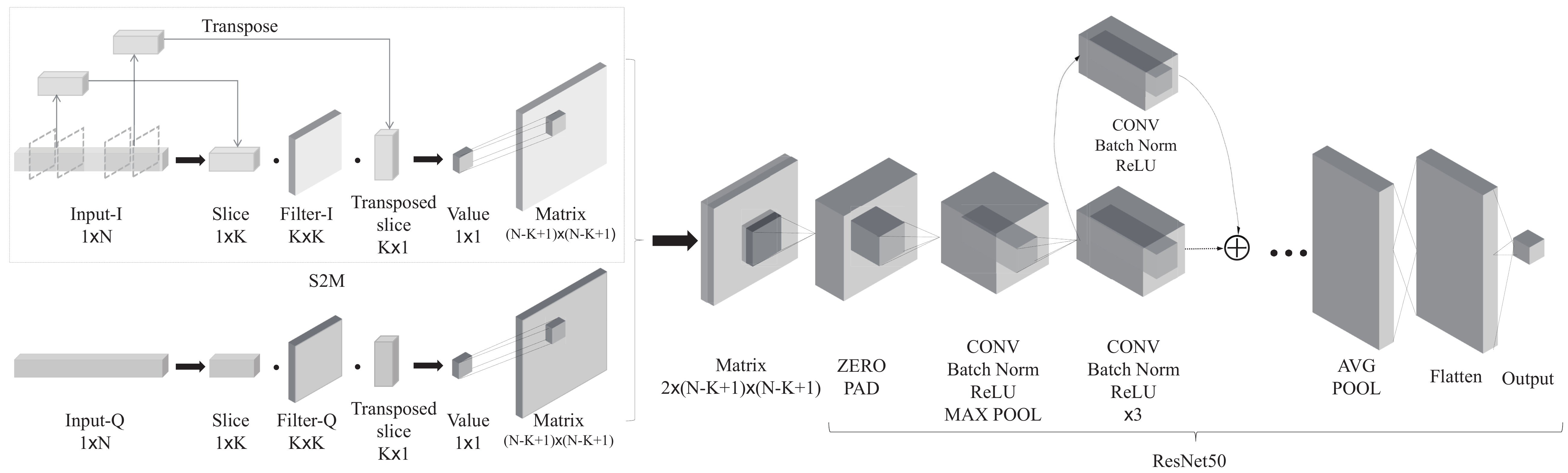}
\caption{The framework of SigNet. A signal-to-matrix operation S2M is implemented by matrix multiplication of signal slices and trainable filters and then a typical CNN (ResNet50) is used for classification. They are trained together by neural network optimization algorithms.}
\label{Fig:overall}
\end{figure*}

\section{Method}
\label{sec:method}
In this section, we introduce our method in detail. Specifically, for the IQ signal, we first design a series of parameter-trainable matrix operations, transform them into matrices, and then classify them with some typical deep learning model. The overall framework of SigNet is shown in Fig.~\ref{Fig:overall}. We mainly focus on the trainable signal-to-matrix method and the work-flow of the entire framework. Finally, based on our basic model SigNet, we also introduce the accelerated version, i.e., SigNet2.0, by integrating 1D and 2D convolution operators, whose structure is shown in Fig.~\ref{Fig:Mix}.

\subsection{S2M operator}
The received radio signal can be seen as a function of time, it is usually represented by its in-phase and the quadrature (IQ) components as
\begin{eqnarray}
x(t) = x_I(t) + jx_Q(t).
\end{eqnarray}
The real and imaginary parts of $x(t)$ represent the I and Q components, respectively. Usually, the signal is sampled at a certain frequency, and the length of each signal sample is fixed to form a dataset.

For an IQ signal sample, we first treat I channel and Q channel separately, and transfer them into matrices by using different signal-to-matrix (denoted by S2M briefly) operators.
Specifically, to facilitate the implementation of the algorithm, an IQ signal can be represented by:
\begin{eqnarray}
\mathbf{I}&=&[i_1, i_2, i_3, ..., i_N], \\
\mathbf{Q}&=&[q_1, q_2, q_3, ..., q_N],
\end{eqnarray}
where $N$ is the signal length of each channel. Without loss of any generalization, here we focus on I channel, and the operation for Q channel is exactly the same as I channel. Firstly, we use a sliding window to sample the signal of I channel, where the length of the window, denoted by $k$, is adjustable. By the sliding window, we get a slice of signal with length $k$. The window moves from $[i_1,i_2,...,i_k]$ to the end of the signal $[i_{N-k+1},i_{N-k+2},...,i_N]$, with the default stride set to $h=1$. From the start to the end, putting all the slices together, we thus get a slice matrix as follows:

\begin{equation}
\mathbf{S}_I=\left[
 \begin{matrix}
   i_1 & i_2 & \cdots & i_k \\
   i_2 & i_3 & \cdots & i_{k+1} \\
   \vdots & \vdots & \ddots & \vdots \\
   i_{N-k+1} & i_{N-k+2} & \cdots & i_N
  \end{matrix}
  \right],\label{EQ:SI}
\end{equation}
with the matrix size $(N-k+1)\times{k}$, meaning there are totally $N-k+1$ slices of length $k$.

Now, we define a filter $F_I$ as our trainable operator, which is represented by a square matrix:
\begin{equation}
\mathbf{F}_I=\left[
 \begin{matrix}
   \alpha_{11} & \alpha_{12} & \cdots & \alpha_{1k} \\
   \alpha_{21} & \alpha_{22} & \cdots & \alpha_{2k} \\
   \vdots & \vdots & \ddots & \vdots \\
   \alpha_{k1} & \alpha_{k2} & \cdots & \alpha_{kk}
  \end{matrix}
  \right],\label{EQ:FI}
\end{equation}
with the matrix size $k\times{k}$. The values of elements in $F_I$ are randomly set by standard Gaussian distribution initially, and can be update by optimization algorithms in the training process of the whole deep learning framework.

Based on the slicing matrix $S_I$ and the filter $F_I$, we can transform a signal of I channel into a feature matrix by
\begin{equation}
\mathbf{M}_I= \mathbf{S}_I \cdot \mathbf{F}_I \cdot \mathbf{S}_I^\mathrm{T},
\end{equation}\label{EQ:MI}
with the matrix size $(N-k+1)\times{(N-k+1)}$.

Note that the elements in matrix $M_I$ can be considered as a kind of dot product of two signal slices at different positions with the operator $F_I$:
\begin{equation}
m_I(h,l) = [i_h,i_{h+1},\cdots,i_{h+k}] \cdot \mathbf{F}_I \cdot {[i_l,i_{l+1},\cdots,i_{l+k}]}^\mathrm{T},  \label{EQ:mpq}
\end{equation}
where $m_I(h,l)$ represents the element of matrix $M_I$ at position $(h, l)$. Since dot product of two vectors generally represents the correlation between them, feature matrix $M_I$ can capture the local auto-correlation of the signal under appropriate transformation, to certain extent.

Similarly, we can get the feature matrix of the Q-channel signal:
\begin{equation}
\mathbf{M}_Q = \mathbf{S}_Q \cdot \mathbf{F}_Q \cdot \mathbf{S}_Q^\mathrm{T}, \label{EQ:MQ}
\end{equation}
where $S_Q$, $F_Q$ represent the signal slice matrix and filter matrix of the Q-channel signal, denoted as follows, respectively.
\begin{equation}
\mathbf{S}_Q=\left[
 \begin{matrix}
   q_1 & q_2 & \cdots & q_k \\
   q_2 & q_3 & \cdots & q_{k+1} \\
   \vdots & \vdots & \ddots & \vdots \\
   q_{N-k+1} & q_{N-k+2} & \cdots & q_N
  \end{matrix}
  \right]\label{EQ:SQ},
\end{equation}
\begin{equation}
\mathbf{F}_Q=\left[
 \begin{matrix}
   \beta_{11} & \beta_{12} & \cdots & \beta_{1k} \\
   \beta_{21} & \beta_{22} & \cdots & \beta_{2k} \\
   \vdots & \vdots & \ddots & \vdots \\
   \beta_{k1} & \beta_{k2} & \cdots & \beta_{kk}
  \end{matrix}
  \right]\label{EQ:FQ}
 \end{equation}

In this way, we convert the IQ signal into two matrices $M_I$ and $M_Q$, which can be summarized as Algorithm~\ref{alg:1}.

\begin{algorithm}[!h]
    \renewcommand{\algorithmicrequire}{\textbf{Input:}}
    \renewcommand{\algorithmicensure}{\textbf{Output:}}
    \caption{S2M operation}
    \label{alg:1}
    \begin{algorithmic}[1]
        \REQUIRE
        IQ signal: $\mathbf{I}=[i_1, i_2, ..., i_N]$, $\mathbf{Q}=[q_1, q_2, ..., q_N]$;\\
        Parameters: filter size $k$.
        \ENSURE
        Matrices $\mathbf{M}_I$ and $\mathbf{M}_Q$.
        \STATE Get slice matrices $\mathbf{S}_I$ and $\mathbf{S}_Q$ by Eq.~(\ref{EQ:SI}) and Eq.~(\ref{EQ:SQ}), respectively.\\
        \STATE Get trainable filters $\mathbf{F}_I$ and $\mathbf{F}_Q$:\\
        If initialization: $\mathbf{F}_I\sim\mathcal N(0,1)$, $\mathbf{F}_Q\sim\mathcal N(0,1)$; \\
        If training: $\mathbf{F}_I=\mathbf{F}_I-\eta\nabla L(\mathbf{F}_I)$, $\mathbf{F}_Q=\mathbf{F}_Q-\eta\nabla L(\mathbf{F}_Q)$,
        where $L$ is the classification loss of entire deep learning framework, and $\eta$ is the learning rate.
        \STATE Obtain the feature matrices: \\
        $\mathbf{M}_I = \mathbf{S}_I \cdot \mathbf{F}_I \cdot \mathbf{S}_I^\mathrm{T}$; \\
        $\mathbf{M}_Q = \mathbf{S}_Q \cdot \mathbf{F}_Q \cdot \mathbf{S}_Q^\mathrm{T}$.
        \STATE Return $\mathbf{M}_I$, $\mathbf{M}_Q$.
    \end{algorithmic}
\end{algorithm}

\subsection{Work-flow of SigNet}
Through S2M operator, we can get two feature matrices $M_I$ and $M_Q$ corresponding to I and Q channels, respectively. Then, to do signal classification, we concatenate the two matrices in the channel dimension to form a two-channel image, which is then fed into a CNN for training and classification, as shown in Fig~\ref{Fig:overall}. Here, the CNN structure could be any typical one, such as Alexnet, VGG and ResNet etc.

The training of the overall deep learning framework can be divided into two parts: the first is training for the signal filters ($\mathbf{F}_I$ and $\mathbf{F}_Q$) and the second is training for CNN. The filter training is to find better matrix representations of signals, and the CNN training is to find higher feature representations from these matrices. Note that the whole framework is end-to-end, while the filters and CNN can be either trained together or alternately. Also, we can set different training parameters for them, such as different learning rate or different weight decay proposals. In this paper, for convenience, we train them together with the same training parameters.

In this way, we can make full use of the capacity of CNN in processing images (or matrices) without destroy the inherent structure of the signals. Simulation results confirm the performance of our method.

\subsection{Principal and advantages}
The core of SigNet is to develop a new operator for signal processing, which can be adaptively trained when naturally integrated into the framework of CNN. And the motivation of using a sliding window and a trainable filter to transfer 1D signal into 2D matrix is that we believe, with the end-to-end training, the trained operator can better convert the signal into a two-dimensional representation. Just as the automatic image classification of CNN has surpassed the image classification of artificially designed features, an automatic image representation of signal may be better than deterministic signal-analysis-based and specific-mathematical-formulas-based approaches.

\begin{figure*} [!t]
\centering
\includegraphics[width=0.99\textwidth]{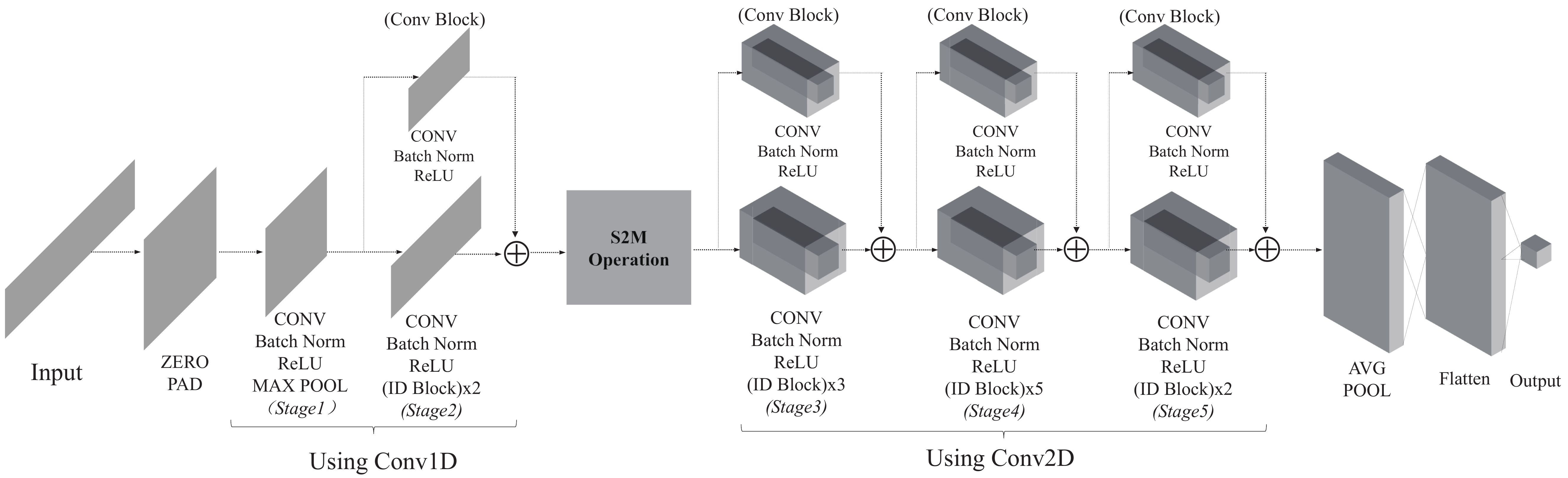}
\caption{One kind of structure of SigNet2.0, which is based on ResNet50. We try to use 1D convolution to replace some of the 2D convolution of SigNet to accelerate the model. Here we choose replacing 1D convolution with 2D convolution on stage1 and stage2, for it's highest classification accuracy.}
\label{Fig:Mix}
\end{figure*}

Signals are typical time series, with the key information to distinguish each other hidden in the correlation between segments. The information at each point is simple, while that of the whole signal is complex. We thus first get continuous slices of signal by a sliding window with length $k$, denoted by $x_h, h=1,2,\cdots,x_{N-k+1}$, with $x_h=[i_h,i_{h+1},\cdots,i_{h+k-1}]$ for I channel and $x_h=[q_h,q_{h+1},\cdots,q_{h+k-1}]$ for Q channel; then we calculate the correlation between each pair of them after a linear mapping function $g(x_h)=G\cdot{x_h}'$, with $G$ the square matrix with size $k$, to get the feature matrix:
\begin{equation}
m_{hl} = g(x_h)^\mathrm{T}g(x_l)=x_hG^\mathrm{T}Gx_l', \label{EQ:mhl}
\end{equation}
which will be equivalent to Eq.~(\ref{EQ:mpq}) when we simply set $F_I=G^\mathrm{T}G$. Interestingly, when $F_I$ or $G$ is set to identity matrix, we will get
\begin{equation}
m_{hl} = x_hx_l',
\end{equation}
which degenerates to Gram matrix and characterizes the correlation between two slices of the signal at different positions.

In this sense, the feature matrix of S2M is actually a transformed correlation matrix of all the signal slices (or namely a transformed local auto-correlation matrix for the whole signal). 
Note that, in a previous work~\cite{wang2016imaging}, Wang et al. used the Gramian Angular Fields to convert time series into different types of images for signal classification. The main difference is that we don't use the Angular Field to transfer the values to some angles, instead, we use a trainable filter to transform the representational space of signal slices first and then calculate their inner products. The obtained feature matrix could be considered as a generalized version of Gram matrix based on signal slices. And SigNet thus has the following advantages: First, through training with the CNN, a more suitable representational space can be automatically found for classification; Second, the original Gram matrix is a symmetric matrix, which is big but of huge information redundancy. By breaking such symmetry, more information could be utilized, since it could be different for exchanging two slices when multiply by a transformation matrix between them.

Compared with those methods that directly utilize signals, such as reshape, narrow 2D-CNN and 1D-CNN, our SigNet makes full use of feature extraction capabilities of CNNs without destroying the structural information of original signals, leading to the fluent learning of the whole framework. SigNet is such a flexible framework, one can adjust the size of filter to find the information-unit on a longer or shorter slice, and change the stride to reduce the input size when signal is especially long. We will continue to explore these hyper-parameter sensitivities and researchers can also adjust them according to their own tasks to get a better performance.

\subsection{Acceleration and upgrade by integrating 1D convolution}

SigNet converts the original signal sequence into a matrix, which can enrich the hidden features of the signal. However, this conversion greatly increases the size of the input data, and the model will occupy a lot of computing resources, especially for a long input sequence. On the other hand, 1D-CNN is naturally adopted to classify signals, 
which has the following advantages: Firstly, it directly uses the signal sequence without destroying its inherent structure; Secondly, it is fast and effective. O’Shea et al.~\cite{8267032} designed a 1D-CNN model and made it suitable for small radio signal classification tasks. Therefore, here, we consider to integrate 1D-CNN into our framework so as the accelerate and upgrade our model.

We try to use 1D convolution operators to replace some of the 2D convolution of SigNet to accelerate the framework, and the specific number of replacement layers depends on the structure of the CNN. As shown in Fig.~\ref{Fig:Mix}, for example, we use ResNet50~\cite{he2016deep} as the basic CNN structure for classification, which consists of 5 stages each with a convolution block and several identity blocks, each convolution block has 3 convolution layers and each identity block also has 3 convolution layers. For the original SigNet, the signal sequence will be converted into a matrix first  by S2M and then input into ResNet50 for classification, but now we consider to use 1D convolution first to reduce the length of the signal sequence and thus reduce the size of the transformed matrix. Begin from the input layer, we replace some of 2D convolutional layers of ResNet50 with 1D convolutional layers (we repalced stage1 and stage2, actually), these 1D convolution structure can be seen as the part of 1D ResNet50, it will output the feature sequence of the input signal, which is much shorter than the input signal but could contain most of the useful information. Then, an S2M operation will be deployed to convert the feature sequence to a matrix and the rest of the structure of ResNet50 will do the final classification. We name the upgraded model SigNet2.0 which mixes the 1D structure and 2D structure of the CNN for signal classification. Simulation results show that SigNet2.0.0 can greatly improve the efficiency of model training without loss of classification accuracy. Impressively, when the ratio of 1D structure and 2D structure is set appropriately, it can even improve the classification accuracy of the model to a certain extent.

\section{Simulations}
\label{sec:experiments}

In this section, we perform the simulations to validate the effectiveness of our models. All the deep learning models are trained on NVIDIA Tesla V100-PCIE. 

\subsection{Modulation dataset}
We evaluate our methods on the following two datasets for signal modulation classification.

\begin{itemize}
    \item \textbf{RML2016.10a} This dataset is public for signal modulation classification~\cite{o2016radio, o2016convolutional}. It uses GNU Radio~\cite{BlossomGNU} to synthesize IQ signal samples. The dataset contains 11 modulation categories consisting of BPSK, QPSK, 8PSK, 16QAM, 64QAM, BFSK, CPFSK, and PAM4 for digital modulations, and WB-FM, AM-SSB, and AM-DSB for analog modulations. The signal to noise ratio (SNR) ranges from -20dB to 18dB, the length of each sample is 128, and the total size of the dataset is 220,000. We divide the dataset into training set, validation set and test set according to the ratio of 6:2:2, the size of training set is 132000 (600 samples per modulation type of each SNR), the size of validation set and test set are both 44000 (200 samples per modulation type of each SNR).
    \item \textbf{Sig2019-12} This dataset contains longer signals simulated by ourselves. The simulation considers several nonideal effects of a real communication system, including carrier phase, pulse shaping, frequency offsets and noise. It contains 12 modulation categories consisting of BPSK, QPSK, 8PSK, OQPSK, 2FSK, 4FSK, 8FSK, 16QAM, 32QAM, 64QAM, 4PAM, and 8PAM. The original data is generated in a random manner, so as to guarantee equal probability of transmitted symbols. The pulse shaping filter is a raised cosine filter, and the roll-off factor is a random value within the range 0.2 to 0.7. The phase offset is randomly selected within the range $-\pi$ to $\pi$, and the carrier frequency offset (normalized to the sampling frequency) is randomly selected within the range -0.1 to 0.1. The SNR of each modulation type is evenly distributed from -20dB to 30dB. Each data sample contains 64 symbols, and the oversampling rate is 8, so the number of sampling points for each sample is 512. The total size of the dataset is 468000 (1500 samples per modulation type of each SNR), and we divide dataset into training set, validation set and test set according to the ratio of 6:2:2, just the same as the division strategy of RML2016.10a dataset.
\end{itemize}

In the following signal modulation classification tasks, all simulations are based on these two datasets, for RML2016.10a, because all examples were scaled to unit energy~\cite{o2016radio}, we use the raw data as input; for Sig2019-12, we use the maximum-minimum normalization to normalize each sample, which can be formulated as $x'_i=(x_i-min)/(max-min)\times 2-1$, where $x_i$ is the original value of each sampling point in a signal channel (I or Q), $max$ and $min$ are the maximum and minimum values of the sampling points in a signal channel, respectively. Without special instructions, we use entire training set for training, entire validation set for validation and entire test set for testing.

\subsection{Models' setup and baselines}
\label{setup}
To do signal classification, we set our SigNet with 2 filters of shape $3\times3$ and 1 stride in S2M operation, so it can transfer an IQ signal to a 2-channel matrix, each 128-length signal will be converted into a $126\times126\times2$ image and each 512-length signal will be converted into a $510\times510\times2$ image. Such image then will be imported into a default ResNet50~\cite{he2016deep} model implemented by Pytorch~\cite{paszke2017automatic}.

There are several kinds of baselines to be compared so as to validate the effectiveness of our models. The first kind is to directly transfer the signal as 2D input, specifically, we reshape each signal sample into a square matrix, a 128 (or 512) length IQ signal can be concatenated into a 256 (1024) length sequence, and then it can be further reshaped as a $16\times16$ (or $32\times32$) matrix, which is then fed into the ResNet50. In the field of time-series analysis, there are also methods to convert a time-series to a matrix first using some theories, such as GAF and MTF, and then connect to typical CNN models to get classification results. The brief introduction of these methods are provided in Sec.~\ref{sec:method}, and there is also a method that uses signal constellation diagram and deep learning models for classification\cite{peng2018modulation}. Here we use these methods as the second kind of baselines which mainly use some specific mathematical method to convert signal to matrix. There are also several existing researches that using deep learning methods to do signal classification, such as narrow 2D-CNN~\cite{o2016convolutional}, 1D-ResNet~\cite{8267032}, MCLDNN~\cite{xu2020spatiotemporal}, and LSTM~\cite{rajendran2018deep}. We use them as the third kind of baselines. We re-implemented the above-mentioned methods in the same programming environment based on Python3, implemented all the deep learning models for simulation based on the deep learning framework Pytorch, and performed simulations based on our datasets division. For dataset RML2016.10a, the architecture description of these models are shown in Table~\ref{table0}, the layers of the model from input to output are described in order. And for dataset SigNet2019-12, there is no changes but the 
number of output neurons will be 12.

\begin{table}[!h]
    \small
    \caption{Architecture description of different models for RML2016.10a dataset.}
    \label{table0}
    \centering
    \begin{tabular}{c|c}
        \toprule
       Models & Architecture description \tabularnewline
        \hline
        1D-ResNet & \makecell[c]{6 ResidualStack each contains 5 Conv1D \\ of kernel size 3 and pool size 2, Dense(128), \\Dropout(0.3), Dense(11)} \tabularnewline
         \hline
        2D-CNN & \makecell[c]{Conv2D(256x1x3), ReLU, Dropout(0.5), \\ Conv2D(80x2x3), ReLU, Dropout(0.5), Dense(256), \\ReLU, Dropout(0.5), Dense(11), softmax(11)} \tabularnewline
         \hline
        LSTM & LSTM(128), LSTM(64), Dense(11) \tabularnewline
         \hline
        MCLDNN & \makecell[c]{Conv1D(50x7) concats with (Conv1D(100x7), \\Conv1D(50x7)), ReLU, Conv1D(100x5), \\LSTM(128), LSTM(128), Dense(128), \\Dropout(0.5), Dense(128), Dropout(0.5), \\Dense(11)} \tabularnewline
         \hline
        SigNet & S2M with a typical 2D ResNet50 \tabularnewline
         \hline
        SigNet2.0 & S2M with a 1D-2D Fusion ResNet50 \tabularnewline
        \bottomrule
    \end{tabular}
\end{table}

For each model training, firstly, we set an epoch number to ensure the model’s testing accuracy reaches convergence after training. Then, after each epoch of training, the model will be validated on the validation set. Finally, the epoch of training with the highest validation accuracy will be the last epoch of training. We select all the models with the highest validation accuracy, save them and use them for testing and performance comparison. We use a warm-up cosine annealing strategy for learning rate decay with an Adam optimizer, and for each model, we used the initial learning rate 0.001 and 0.1 for RML2016.10a and Sig2019-12 datasets, respectively, to ensure that it can be effectively trained. We set batch size of 128 and 32 for each model with RML2016.10a dataset and Sig2019-12 dataset, respectively.

\subsection{Effectiveness of S2M operator}
We first compare the classification results obtained by SigNet with those obtained by the first and second kinds of baselines, by simply reshaping input (Reshape) or using specific mathematical input conversion (GAF, MTF, and constellation diagram), to validate the effectiveness of our trainable S2M operator. 

\begin{figure*} [!t]
\centering
\includegraphics[width=0.99\textwidth]{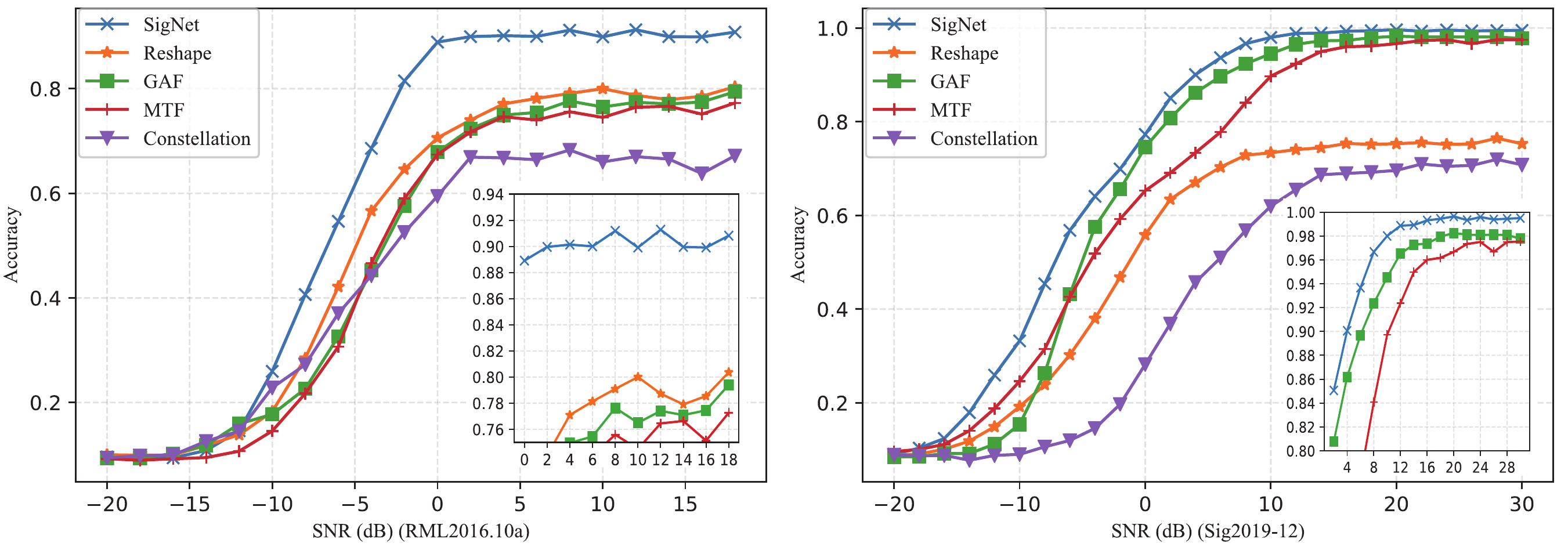}
\caption{The comparison results of SigNet/SigNet2.0 and other signal-to-matrix methods. All the methods do the signal classification by converting signal to matrix first and then using a typical ResNet50 for classification.}
\label{Fig:result1}
\end{figure*}

We train the models by using the whole training set, and then give the classification accuracy on test samples of each SNR. The results are shown in Fig.~\ref{Fig:result1}, where we can see that, on both datasets, SigNet has a higher accuracy than the others. Some signal conversion methods, such as GAF and MTF, perform quite differently on different datasets, which may suggest that these methods lack good generalization ability. All of these indicate that trainable S2M operators can indeed match better with CNN classifier, i.e., it can adaptively capture the better features of signals for classification.

\subsection{Comparison with other deep learning models}
\begin{figure*} [!t]
\centering
\includegraphics[width=0.99\textwidth]{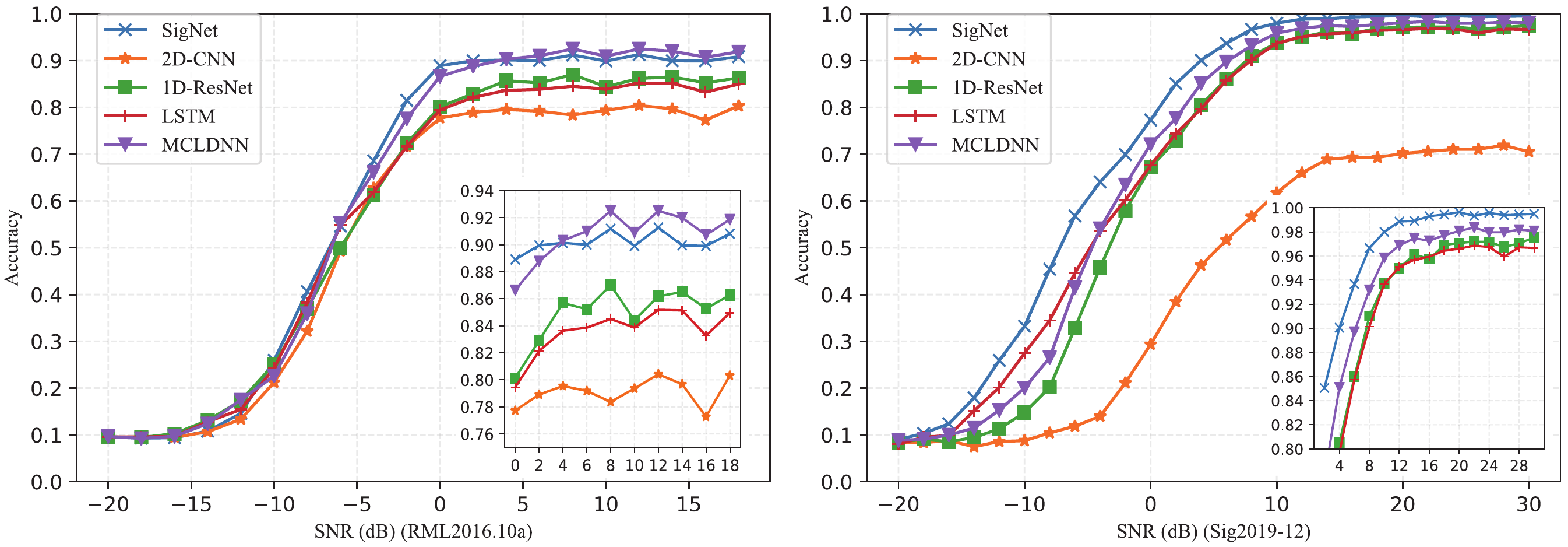}
\caption{The comparison results of SigNet/SigNet2.0 and other deep learning models.}
\label{Fig:state-of-art compare}
\end{figure*}
In this part, we compare SigNet with several state-of-the-art deep learning models applying in signal modulation classification. In particular, we choose three CNN models, narrow 2D-CNN~\cite{o2016convolutional}, 1D-ResNet~\cite{8267032} and  MCLDNN~\cite{xu2020spatiotemporal}, and a 2-layer LSTM model~\cite{rajendran2018deep} as the baselines. The first three are the most advanced CNN models for signal modulation classification, while the last one is a commonly used sequence deep learning model. The results are shown in Fig.~\ref{Fig:state-of-art compare}, as we can see, for dataset RML2016.10a, SigNet performs significantly better than other models, except the state-of-the-art model MCLDNN, the performance of these two models is very close (the accuracy of SigNet on the entire test set is 61.35\%, while that of MCLDNN is 61.16\%). And for dataset Sig2019-12, SigNet performs significantly better than all the other models, especilly when the SNR is below 0dB (the accuracy of SigNet on the entire test set is 72.69\%, while the second best model MCLDNN is 67.52\%). These results indicate that our model has a better generalization than MCLDNN and has a stable performance on more datasets.

\begin{figure*}[htbp] 
\begin{minipage}[t]{0.99\linewidth} 
\centering 
\includegraphics[width=0.99\textwidth]{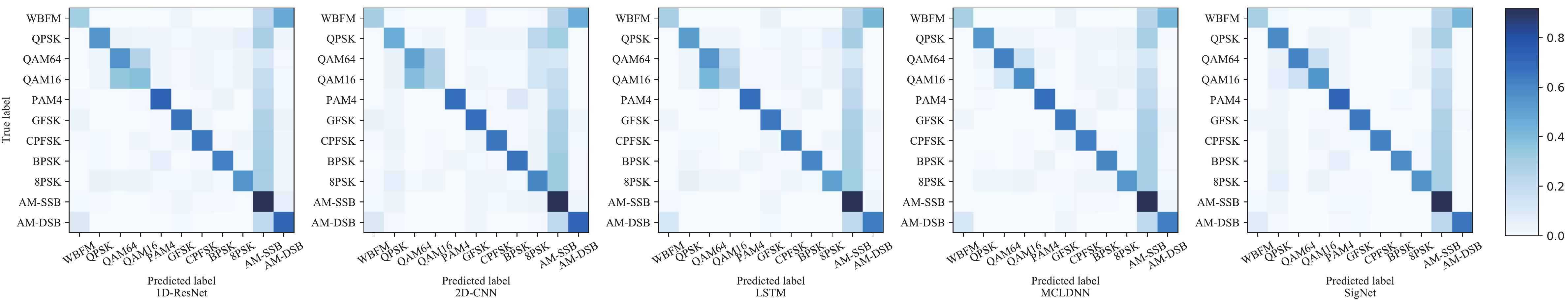} 
\end{minipage}%
\vfill
\begin{minipage}[t]{0.99\linewidth} 
\centering 
\includegraphics[width=0.99\textwidth]{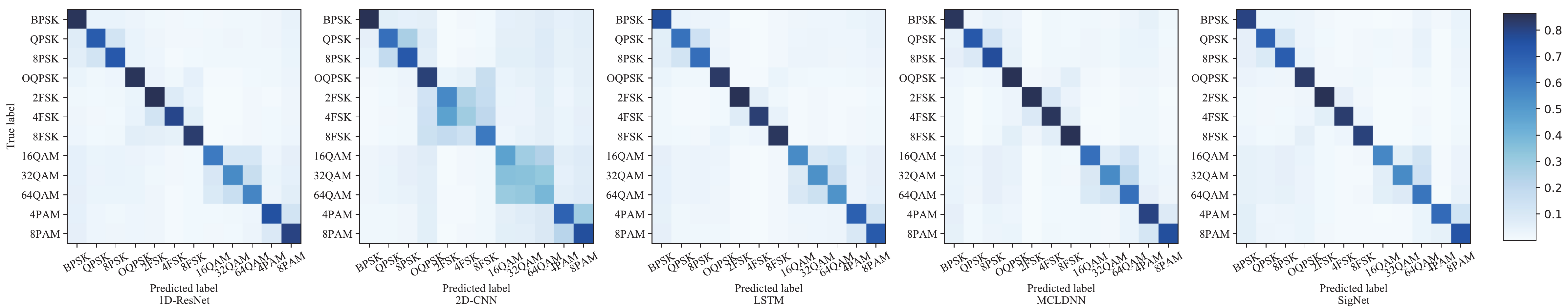} 
\end{minipage} 
\caption{The confusion matrices of five deep learning models, including 1D-ResNet, 2D-CNN, LSTM, MCLDNN and SigNet, on RML2016.10a (top) and Sig2019-12 (bottom).} 
\label{Fig:confusion}
\end{figure*}

To compare the classification performance of these models more comprehensively, we visualize the confusion matrices of the entire test sets in Fig.~\ref{Fig:confusion}, for the two datasets RML2016.10a and Sig2019-12, respectively. As we can see, on the dataset RML2016.10a, 1D-ResNet is more likely to misclassify the modulation category WBFM as AM-DSB; both 1D-ResNet and 2D-CNN are more likely to confuse QAM64 and QAM16, while it seems that both LSTM and SigNet are able to reduce such confusion to certain extent. All the models may misclassify WBFM as AM-SSB or AM-DSB and misclassify all modulation categories as AM-SSB. On the dataset Sig2019-12, 2D-CNN behaves the worst in many categories, especially for 2FSK and 4FSK, while the others perform well. The diagonal of SigNet and MCLDNN's confusion matrices are more distinct than the others, especially for OQPSK, 2FSK, 4FSK and 8FSK. LSTM seems close to SigNet, but the latter has a much higher training efficiency, i.e., comparing with LSTM, SigNet saves 49.58\% and 32.40\% training time on the two datasets RML2016.10a and Sig2019-12, respectively.

\subsection{The performance of SigNet2.0}

To further accelerate our model, we replace some of the 2D-convolutional layers with 1D-convolutional layers, and found that it can slightly increase the classification accuracy when the ratio of 1D-convolutional layers and 2D-convolutional layers is set appropriately. We tried to replace different numbers of 2D-convolutional layers with 1D-convolutional layers and tested the accuracy on the same dataset. For the ResNet50, we found that when the 2D-convolutional layers of stage1 and stage2 were replaced with 1D-convolutional layers (as shown in Fig.~\ref{Fig:Mix}), the model has the highest accuracy with a certain acceleration effect, and we call this model SigNet2.0. Comparing with SigNet, SigNet2.0 improves 0.95\% and 0.18\% accuracy, meanwhile saving 30.65\% and 12.46\% training time, on RML2016.10a and Sig2019-12 datasets, respectively.

\begin{table*}[!h]
    \small
    \caption{The performances of all the models we used.}
    \label{table1}
    \centering
    \begin{tabular}{ccccccc}
        \toprule
        Dataset & Method & Accuracy & F1 Score & Recall & Training Time & Parameter quantity\tabularnewline
        \hline
        \multirow{6}{*}{RML2016.10a}
        & GAF & 49.43\% & 0.5154 & 0.4943 & 3h18m  & 23.53M \tabularnewline
        & MTF & 48.17\% & 0.5162 & 0.4818 & 2h52m & 23.53M \tabularnewline
        & Constellation & 44.92\% & 0.4507 & 0.4493 & 21m & 23.53M \tabularnewline
        & 1D-ResNet & 57.74\% & 0.5930 & 0.5774 & 26m & 0.1M \tabularnewline
        & 2D-CNN & 54.02\% & 0.5504 & 0.5402 & 21m & 2.71M \tabularnewline
        & LSTM & 57.20\% & 0.5877 & 0.5720 & 1h56m & 0.2M \tabularnewline
        & MCLDNN & 61.16\% & 0.6335 & 0.6117 & 53m & 0.37M \tabularnewline
        & SigNet& 61.35\% & 0.6354 & 0.6135 & 62m & 23.53M \tabularnewline
        & SigNet2.0 & \textbf{62.30\%} & \textbf{0.6487} & \textbf{0.6226} & 43m & 23.52M \tabularnewline
        \hline
        \multirow{6}{*}{Sig2019-12}
        & GAF & 67.31\% & 0.6714 & 0.6732 & 33h24m & 23.53M \tabularnewline
        & MTF & 65.21\% & 0.6561 & 0.6521 & 43h52m & 23.53M \tabularnewline
        & Constellation & 41.78\% & 0.4193 & 0.4178 & 1h15m & 23.53M \tabularnewline
        & 1D-ResNet & 64.47\% & 0.6427 & 0.6447 & 50m & 0.13M\tabularnewline
        & 2D-CNN & 41.95\% & 0.4201 & 0.4196 & 1h15m & 10.57M\tabularnewline
        & LSTM & 66.90\% & 0.6675 & 0.6682 & 19h49m & 0.2M \tabularnewline
        & MCLDNN & 67.52\% & 0.6731 & 0.6738 & 1h42m & 0.37M \tabularnewline
        & SigNet & 72.69\% & 0.7230 & 0.7228 & 15h23m & 23.53M \tabularnewline
        & SigNet2.0 & \textbf{72.87\%} & \textbf{0.7265} & \textbf{0.7228} & 13h28m & 23.52M \tabularnewline
        \bottomrule
    \end{tabular}
\end{table*}

\begin{figure}[htbp] 
\begin{minipage}[t]{0.99\linewidth} 
\centering 
\includegraphics[width=0.99\textwidth]{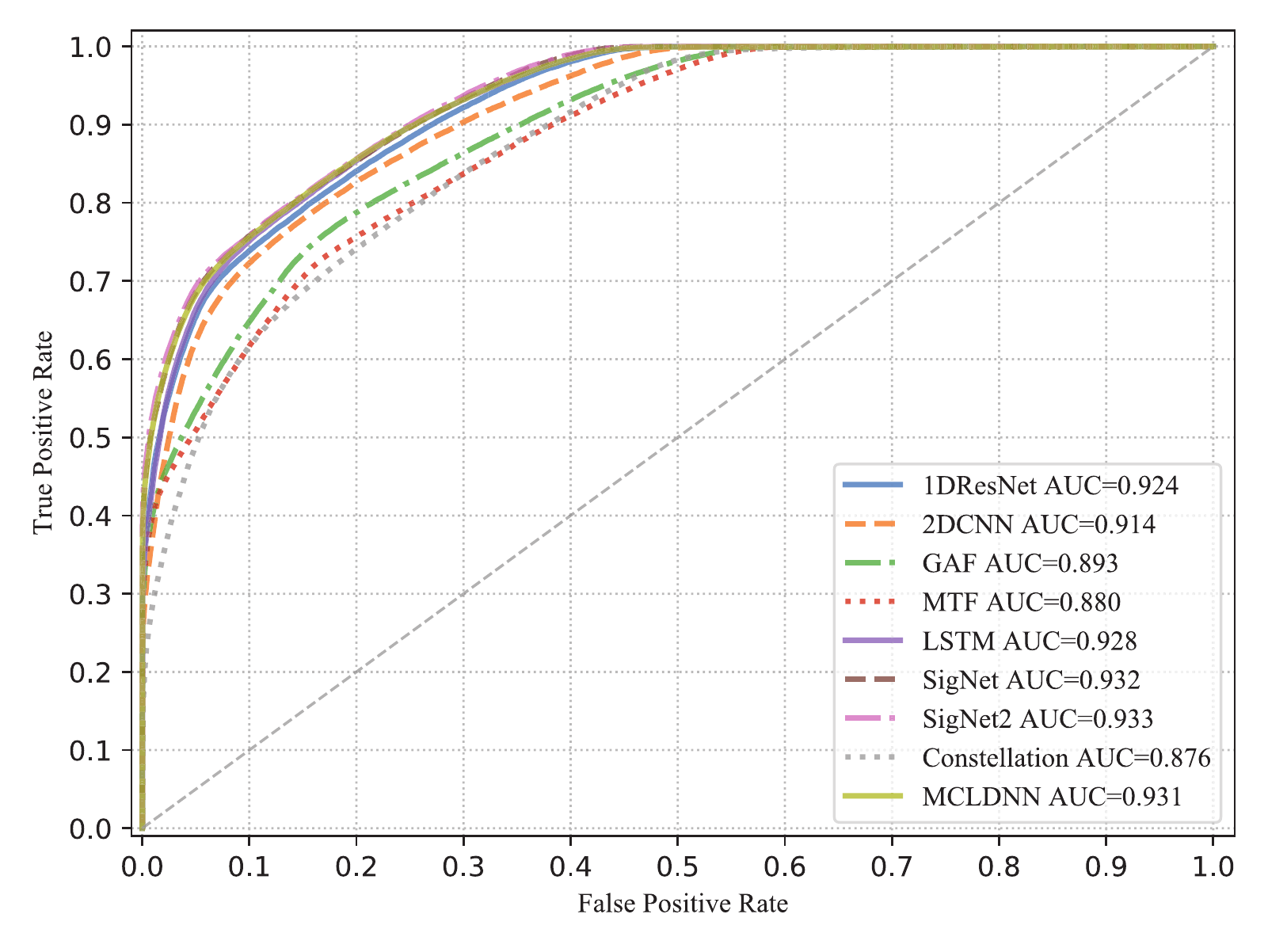} 
\end{minipage}%
\vfill
\begin{minipage}[t]{0.99\linewidth} 
\centering 
\includegraphics[width=0.99\textwidth]{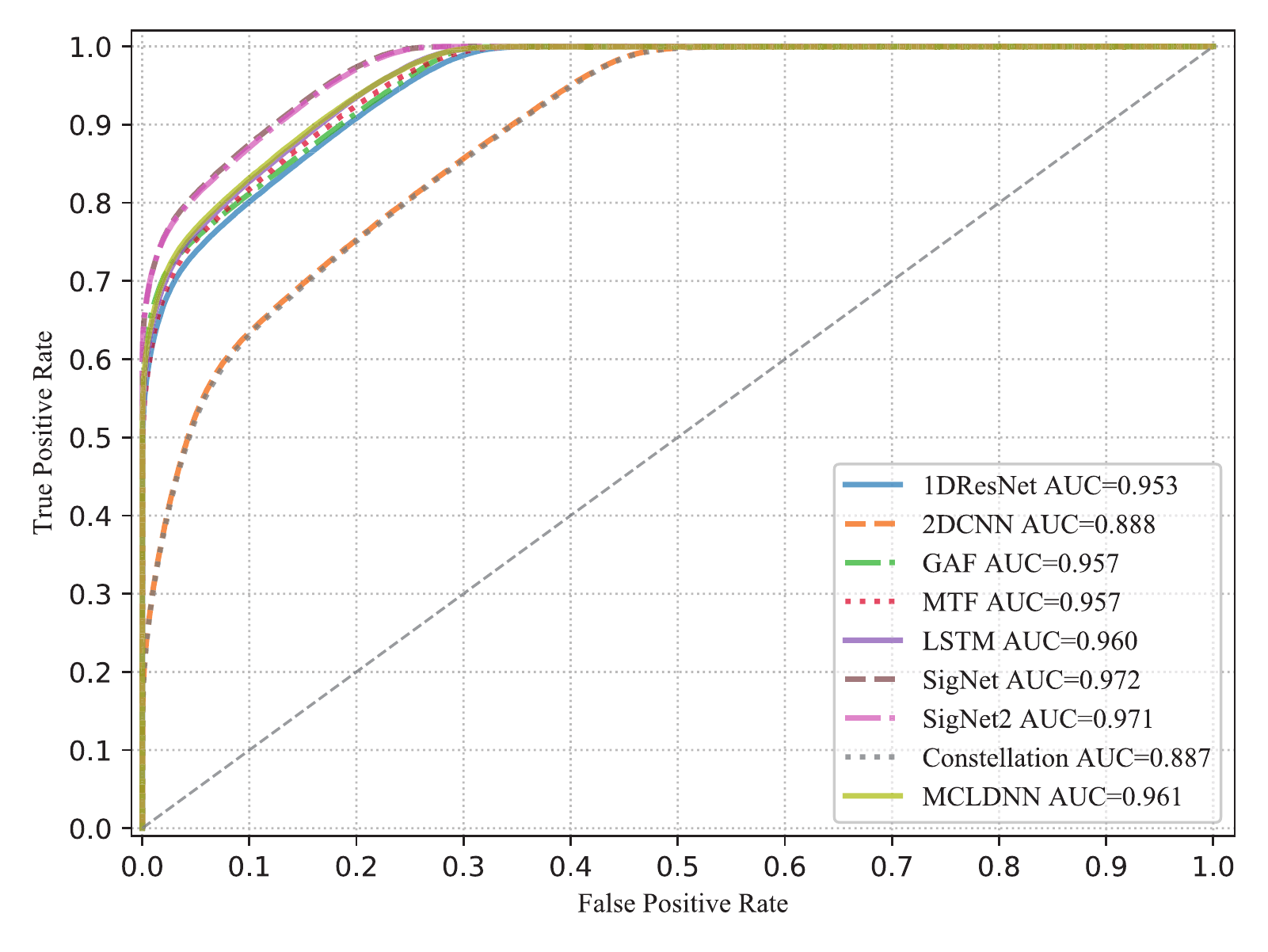} 
\end{minipage} 
\caption{The ROC curves and AUC values of all methods in RML2016.10a test set (top) and Sig2019-12 test set (bottom).}
\label{Fig:ROC}
\end{figure}

We conducted a comprehensive evaluation and comparison of the models mentioned above, including model classification performance, training time and parameter quantity, as shown in TABLE~\ref{table1}. We also used the machine learning toolkit \emph{sklearn} to draw the ROC curves of all methods on the multi-class classification results and calculate the AUC value, as shown in Fig.~\ref{Fig:ROC}. We can see that some models, such as 1D-ResNet and 2D-CNN, are fast but of relatively low classification accuracy; MCLDNN also has 
a sightly lower accuracy than SigNet on the RML2016.10a dataset, although it is lighter and more efficient. On Sig2019-12 dataset, it performs much poorer than our SigNet, indicating that our method has the better generalization ability when the data sample becomes longer. 

Among these methods, GAF, MTF and Constellation have the same amount of trained parameters, but they show great difference in training time. The first reason is that the input size of Constellation is smaller than GAF and MTF and another reason is that, for GAF and MTF, due to the storage space limitation, we use an end-to-end framework to include the processing of signal samples to matrix and subsequent CNN-based classification in one framework, while for Constellation, we first use the counting method to directly convert the entire datasets into density matrices, and then input them into CNNs for classification. So, for the training time measurement, there are two training time measurements for different models. For GAF and MTF, because the processed training set requires huge storage space, we choose to perform preprocessing together with subsequent CNN training, so their training time includes the time of preprocessing and subsequent CNN training. For other models, the training time only includes the training time of the model parameters. Since the training time of GAF and MTF also includes the time of converting signals into matrices, they are more time-consuming than other methods, but their training time is only listed as a reference and is not used for comparison with that of other methods. In general, GAF and MTF methods are not flexible enough in implementation, and the accuracy is worse than that of SigNet. Compared with other models, our SigNet model has more parameters, since it is based on ResNet50, which has a total of 50 convolutional layers. However, our model can be trained efficiently due to the residual structure. Among all the methods, SigNet2.0 performs the best in classification performance, obtaining the state-of-the-art results, reaching a good balance between accuracy and efficiency. 

\subsection{Learning with small samples}
\begin{figure*} [!t]
\centering
\includegraphics[width=0.99\textwidth]{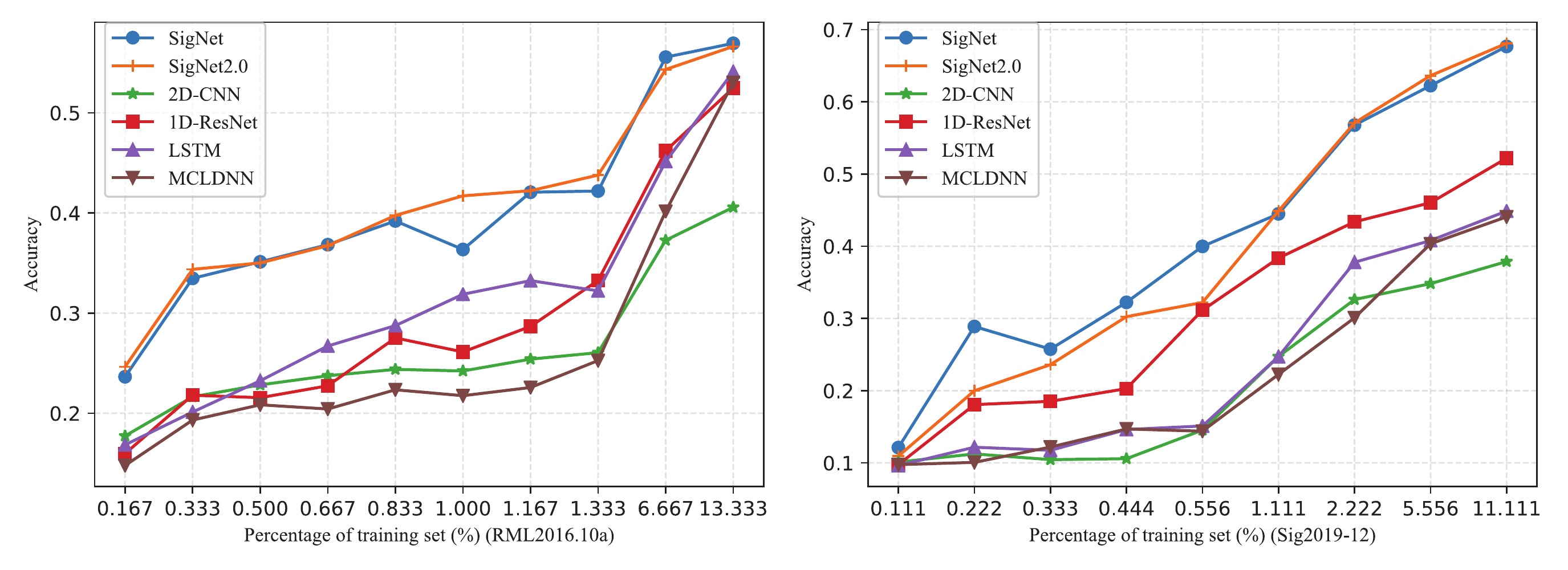}
\caption{The comparison results of our proposed SigNet/SigNet2.0 and those deep learning models as baselines on small sample learning.}
\label{Fig:small_training_set}
\end{figure*}

There are always real situations that we cannot get enough signal samples with labels. In reality, collecting labels for signals may be more expensive than for images, since they are relatively abstract and thus need a large amount of expert labors. To validate the effectiveness of our method in such situations, we also compare SigNet/SigNet2.0 with other state-of-the-art deep learning models by using only a small fraction $\eta$ of training set. In particular, for RML2016.10a, we vary $\eta$ from $1/600$ to $80/600$ (about 0.167\% to 13.333\%), which means that in the least case, each modulation category's each level of SNR only has one sample participates in training, similarly, for Sig2019-12, the $\eta$ is from $1/900$ to $100/900$ (about 0.111\% to 11.111\%), the classification results are show in Fig.~\ref{Fig:small_training_set}. 

We find that, in general, our SigNet and SigNet2.0 can keep relatively high accuracy when a small fraction of training set is adopted: e.g., they achieve 56.94\% and 56.63\%, respectively for RML2016.10a, and achieve 67.67\% and 68.10\%, respectively for Sig2019-12 when less than 15\% training set is adopted; and these numbers are 36.35\% and 41.47\% for RML2016.10a, and 44.49\% and 44.88\% for Sig2019-12 even when 
about 1\% training set is adopted. By comparison, the accuracy of the other deep learning models drops very quickly as the training set gets smaller. Both MCLDNN and LSTM use the features of sequential neural networks (part of the features used by MCLDNN is based on LSTM), which may be the reason for their poor performance in small sample learning on Sig2019-12. Such results are quite impressive by considering that these three models have comparable results when the full training set is adopted. More remarkably, for RML2016.10a, when each modulation category's each level of SNR only has one sample participates in training, SigNet and SigNet2.0 still have the accuracy above 20\% for the both datasets, much higher than the other three models. These results suggest that our SigNet and SigNet2.0 have powerful feature extraction capacity and could be very useful in the situations where the labeled data are difficult to obtain.

\subsection{A brief ablation study of SigNet}
We also conducted a brief ablation study of SigNet's hyper-parameters, including batch size, filter size of S2M and optimizer. Since we use the learning rate decay strategy of warm-up cosine annealing, the ablation study of learning rate has not been taken. The results are shown in Table~\ref{table2}. For dataset RML2016.10a, when the batch size is set to 128, the accuracy of the model is the highest, and when the filter size changes within a certain range (3, 5, 7, 9), the accuracy is almost unaffected. For dataset Sig2019-12, within a certain range, batch size (32, 64, 128) and filter size (3, 5, 7) have little effect on the accuracy of the model.
\begin{table}[!h]
    \small
    \caption{A brief ablation study of SigNet with test accuracy.}
    \label{table2}
    \centering
    \begin{tabular}{ccccc}
        \toprule
        Dataset & BatchSize & FilterSize & Optimizer & Accuracy \tabularnewline
        \hline
        RML2016.10a & 32 & 3x3 & Adam  & 60.60\% \tabularnewline
        RML2016.10a & 64 & 3x3 & Adam  & 60.29\% \tabularnewline
        RML2016.10a & 128 & 3x3 & Adam & 61.35\% \tabularnewline
        RML2016.10a & 256 & 3x3 & Adam & 59.53\% \tabularnewline
        RML2016.10a & 128 & 3x3 & SGD & \textbf{61.44}\% \tabularnewline
        RML2016.10a & 128 & 3x3 & Adagrad & 60.99\% \tabularnewline
        RML2016.10a & 128 & 3x3 & Adadelta & 60.98\% \tabularnewline
        RML2016.10a & 128 & 5x5 & Adam & 60.83\% \tabularnewline
        RML2016.10a & 128 & 7x7 & Adam & 61.18\% \tabularnewline
        RML2016.10a & 128 & 9x9 & Adam & 61.20\% \tabularnewline
        \hline
        Sig2019-12 & 32 & 3x3 & Adam & \textbf{72.69}\% \tabularnewline
        Sig2019-12 & 64 & 3x3 & Adam & 72.14\% \tabularnewline
        Sig2019-12 & 128 & 3x3 & Adam & 71.93\% \tabularnewline
        Sig2019-12 & 32 & 3x3 & SGD & 69.13\% 
        \tabularnewline
        Sig2019-12 & 32 & 3x3 & Adagrad & 70.12\% \tabularnewline
        Sig2019-12 & 32 & 3x3 & Adadelta & 70.01\% \tabularnewline
        Sig2019-12 & 32 & 5x5 & Adam & 72.20\% \tabularnewline
        Sig2019-12 & 32 & 7x7 & Adam & 72.58\% \tabularnewline
        Sig2019-12 & 32 & 9x9 & Adam & 71.98\% \tabularnewline
        \bottomrule
    \end{tabular}
\end{table}

What's more, we also studied the impact of using different ResNet structures on the performance of our framework. We use the hyperparameter settings described in Section~\ref{setup} and replace the subsequent ResNet50 structure of S2M with ResNet18, ResNet34 and ResNet101 respectively, and the classification accuracy is shown in Table~\ref{table3}. We can see that in general, the changes in the ResNet structure do not have much impact on the classification performance of the SigNet framework. For RML2016.10a, ResNet50 is better than other structures, and for Sig2019-12, the simpler structure ResNet34 performs better than others. And no matter which ResNet structure is used with S2M, the signal classification performance is much better than other frameworks that use ResNet, such as GAF, MTF, etc.
\begin{table}[!h]
    \small
    \caption{A brief ablation study of SigNet with test accuracy.}
    \label{table3}
    \centering
    \begin{tabular}{cccc}
        \toprule
        Dataset & Architecture & Accuracy & \makecell[c]{Parameter \\quantity} \tabularnewline
        \hline
        RML2016.10a & S2M+ResNet18 & 60.38\% & 11.17M \tabularnewline
        RML2016.10a & S2M+ResNet34 & 60.48\% & 21.28M \tabularnewline
        RML2016.10a & S2M+ResNet50 & \textbf{61.35\%} & 23.52M \tabularnewline
        RML2016.10a & S2M+ResNet101 & 60.85\% & 42.52M \tabularnewline
        \hline
        Sig2019-12 & S2M+ResNet18 & 72.13\% & 11.17M \tabularnewline
        Sig2019-12 & S2M+ResNet34 & \textbf{72.52\%} &  21.28M \tabularnewline
        Sig2019-12 & S2M+ResNet50 & 72.14\% &  23.52M \tabularnewline
         Sig2019-12 & S2M+ResNet101 & 72.26\% &  42.52M \tabularnewline
        \bottomrule
    \end{tabular}
\end{table}

\section{Explanatory visualization of models}
\label{sec:visual}
In this section, we will visualize the output features of the six deep learning models proposed by other literature, to see how these models distinguish signals of different modulations. 

In particular, we use t-SNE (t-distributed Stochastic Neighbor Embedding)~\cite{maaten2008visualizing} to map the output features extracted by the trained model into a two-dimensional space for visualization. Specifically, we first remove the last softmax layer of the trained deep learning models, and then input the test dataset to obtain the corresponding output features. After that, we use the t-SNE technique to reduce the dimension of the output features for visualization. For the t-SNE technique, the number of iterations is set to 300, the learning rate is 1000, the perplexity is 30, and the momentum changes from 0.5 to 0.8. 

\begin{figure*} [!t]
\centering
\includegraphics[width=0.99\textwidth]{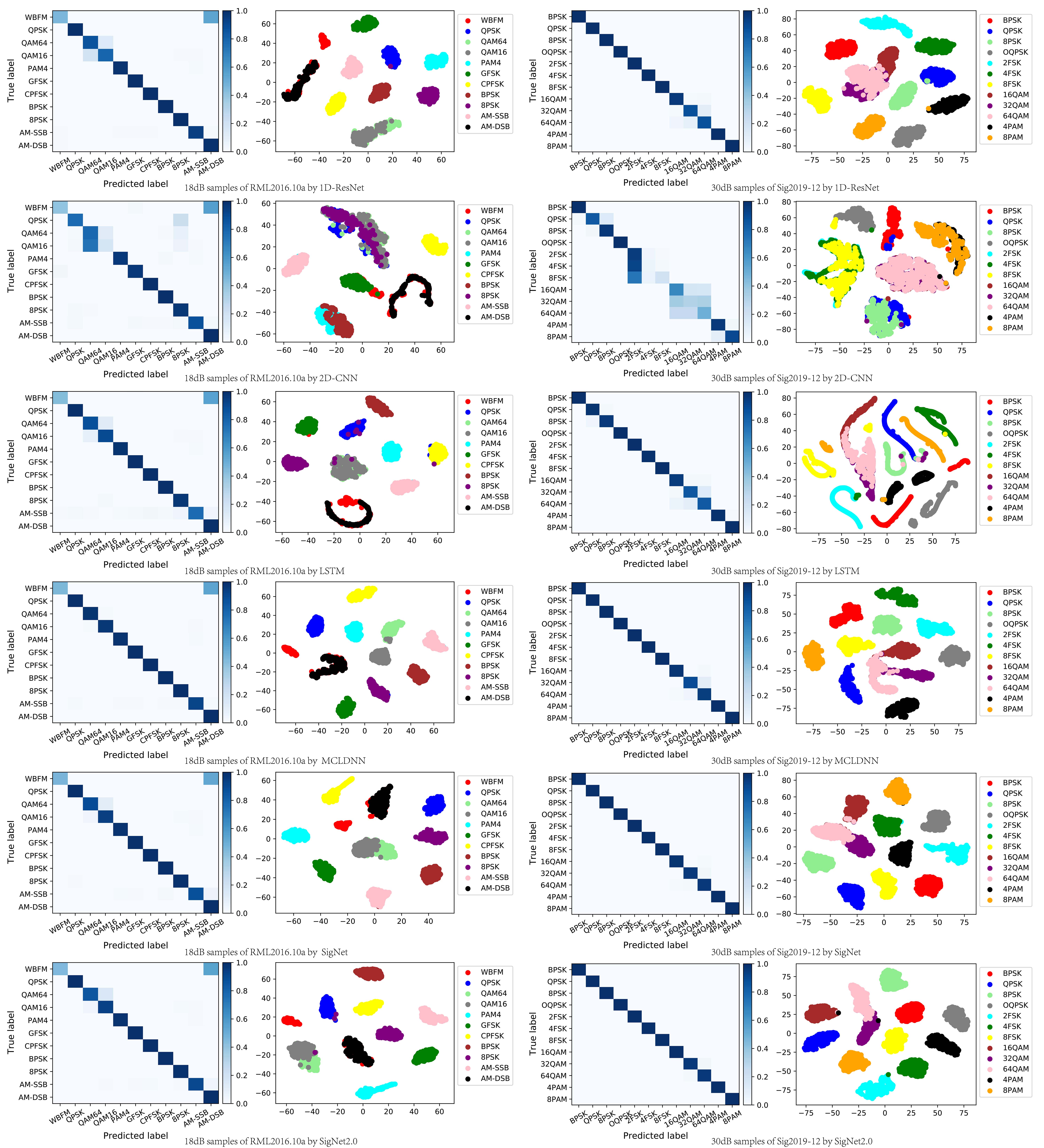}
\caption{The visualization of test samples of the maximum SNR in the two datasets, i.e., 18dB for RML2016.10a (left) and 30dB for Sig2019-12 (right), by using the six deep learning models, including 1D-ResNet, 2D-CNN, LSTM, MCLDNN, SigNet, and SigNet2.0 (from top to bottom).}
\label{Fig:maxdB}
\end{figure*}

We compare SigNet/SigNet2.0 with other deep learning modulation classification models through the output feature visualization. Since the entire test set is too large and complicated for visualization, we only select the data with maximum SNR (18dB for RML2016.10a and 30dB for Sig2019-12), as shown in Fig.~\ref{Fig:maxdB}. Note that we also display the classification results by placing confusion matrices on the left side.

As we can see, the classification results are well reflected in the visualization of the output features. When two categories of data are misclassified with each other by a model, the corresponding data points in t-SNE visualization will also overlap, e.g., the data points of WBFM and AM-DSB in RML2016.10a. It can be observed that the signals of different modulation categories have much clearer boundaries by adopting SigNet/SigNet2.0 and MCLDNN than those by adopting 1D-ResNet, 2D-CNN or LSTM, leading to their much higher classification accuracy, which is consistent with the results presented in TABLE~\ref{table1}.


\section{Conclusion}
\label{sec:conclusion}
In this paper, we proposed a deep learning framework for signal modulation classification, namely SigNet, which first uses a trainable S2M operator to convert the original signal to a square matrix and then use a typical CNN for classification. Moreover, we integrate 1D convolution operators into SigNet, so as to propose SigNet2.0 to further improve the classification accuracy and efficiency. Comprehensive simulation results validate the outstanding performance of our models. Quite impressively, both SigNet and SigNet2.0 give comparable accuracy even when very small training set is adopted, showing their strong feature extraction capacity. Such results indicate that our framework could be especially useful in the situations where labeled signals are difficult to obtain.

In the future, we may try other CNN architectures to further improve the performance of SigNet. Moreover, we will also apply our framework to other kinds of signal datasets to validate its effectiveness in more general cases. We will also consider that single channel sample may contain multiple modulations and further adjust our model to single-input multiple-output to adapt to such a more complex but practical situation.



\bibliography{BIB_TCCN_TPS_20_0171}\

\end{document}